\documentclass[twocolumn,showpacs,aps,prl,superscriptaddress]{revtex4}

\usepackage{graphicx}
\usepackage{dcolumn}
\usepackage{amsmath}
\usepackage{epsfig}
\RequirePackage{xspace}

\usepackage{relsize}
\def\babar{\mbox{\slshape B\kern-0.1em{\smaller A}\kern-0.1em
    B\kern-0.1em{\smaller A\kern-0.2em R}}}

\def\epem       {\ensuremath{e^+e^-}\xspace}

\def\ellell     {\ensuremath{\ell^+ \ell^-}\xspace}

\def\qqbar {\ensuremath{q\overline q}\xspace}

\def\pip   {\ensuremath{\pi^+}\xspace}
\def\pim   {\ensuremath{\pi^-}\xspace}
\def\pipi  {\ensuremath{\pi^+\pi^-}\xspace}
\def\pipm  {\ensuremath{\pi^\pm}\xspace}
\def\pimp  {\ensuremath{\pi^\mp}\xspace}

\def\Kbar  {\kern 0.2em\overline{\kern -0.2em K}{}\xspace}

\def\Kz    {\ensuremath{K^0}\xspace}
\def\Kzb   {\ensuremath{\Kbar^0}\xspace}
\def\KzKzb {\ensuremath{\Kz \kern -0.16em \Kzb}\xspace}
\def\Kp    {\ensuremath{K^+}\xspace}
\def\Km    {\ensuremath{K^-}\xspace}
\def\Kpm   {\ensuremath{K^\pm}\xspace}

\def\KpKm  {\ensuremath{\Kp \kern -0.16em \Km}\xspace}

\def\Kstarz  {\ensuremath{K^{*0}}\xspace}

\def\Dbar    {\kern 0.2em\overline{\kern -0.2em D}{}\xspace}

\def\Dz      {\ensuremath{D^0}\xspace}
\def\Dzb     {\ensuremath{\Dbar^0}\xspace}
\def\DzDzb   {\ensuremath{\Dz {\kern -0.16em \Dzb}}\xspace}
\def\Dp      {\ensuremath{D^+}\xspace}
\def\Dm      {\ensuremath{D^-}\xspace}

\def\DpDm    {\ensuremath{\Dp {\kern -0.16em \Dm}}\xspace}

\def\B       {\ensuremath{B}\xspace}
\def\Bbar    {\kern 0.18em\overline{\kern -0.18em B}{}\xspace}

\def\BB      {\ensuremath{B\Bbar}\xspace} 
\def\Bz      {\ensuremath{B^0}\xspace}
\def\Bzb     {\ensuremath{\Bbar^0}\xspace}
\def\BzBzb   {\ensuremath{\Bz {\kern -0.16em \Bzb}}\xspace}
\def\Bu      {\ensuremath{B^+}\xspace}
\def\Bub     {\ensuremath{B^-}\xspace}
\def\Bp      {\ensuremath{\Bu}\xspace}
\def\Bm      {\ensuremath{\Bub}\xspace}
\def\Bpm     {\ensuremath{B^\pm}\xspace}

\def\BpBm    {\ensuremath{\Bu {\kern -0.16em \Bub}}\xspace}

\def\BorBbar    {\kern 0.18em\optbar{\kern -0.18em B}{}\xspace}
\def\DorDbar    {\kern 0.18em\optbar{\kern -0.18em D}{}\xspace}
\def\KorKbar    {\kern 0.18em\optbar{\kern -0.18em K}{}\xspace}

\def\jpsi     {\ensuremath{{J\mskip -3mu/\mskip -2mu\psi\mskip 2mu}}\xspace}
\def\psitwos  {\ensuremath{\psi{(2S)}}\xspace}

\def\chiczero {\ensuremath{\chi_{c0}}\xspace}

\mathchardef\Upsilon="7107
\def\Y#1S{\ensuremath{\Upsilon{(#1S)}}\xspace}% no space before {...}!

\def\FourS {\Y4S}

\mathchardef\Deltares="7101
\mathchardef\Xi="7104
\mathchardef\Lambda="7103
\mathchardef\Sigma="7106
\mathchardef\Omega="710A

\def\Deltabar{\kern 0.25em\overline{\kern -0.25em \Deltares}{}\xspace}
\def\Lbar{\kern 0.2em\overline{\kern -0.2em\Lambda\kern 0.05em}\kern-0.05em{}\xspace}
\def\Sigbar{\kern 0.2em\overline{\kern -0.2em \Sigma}{}\xspace}
\def\Xibar{\kern 0.2em\overline{\kern -0.2em \Xi}{}\xspace}
\def\Obar{\kern 0.2em\overline{\kern -0.2em \Omega}{}\xspace}
\def\Nbar{\kern 0.2em\overline{\kern -0.2em N}{}\xspace}
\def\Xb{\kern 0.2em\overline{\kern -0.2em X}{}\xspace}

\def\BR         {{\ensuremath{\cal B}\xspace}}

\def\mes        {\mbox{$m_{\rm ES}$}\xspace}

\def\DeltaE     {\mbox{$\Delta E$}\xspace}

\newcommand{\tev}{\ensuremath{\mathrm{\,Te\kern -0.1em V}}\xspace}
\newcommand{\gev}{\ensuremath{\mathrm{\,Ge\kern -0.1em V}}\xspace}
\newcommand{\mev}{\ensuremath{\mathrm{\,Me\kern -0.1em V}}\xspace}
\newcommand{\kev}{\ensuremath{\mathrm{\,ke\kern -0.1em V}}\xspace}
\newcommand{\ev}{\ensuremath{\mathrm{\,e\kern -0.1em V}}\xspace}
\newcommand{\gevc}{\ensuremath{{\mathrm{\,Ge\kern -0.1em V\!/}c}}\xspace}
\newcommand{\mevc}{\ensuremath{{\mathrm{\,Me\kern -0.1em V\!/}c}}\xspace}
\newcommand{\gevcc}{\ensuremath{{\mathrm{\,Ge\kern -0.1em V\!/}c^2}}\xspace}
\newcommand{\mevcc}{\ensuremath{{\mathrm{\,Me\kern -0.1em V\!/}c^2}}\xspace}
\def\invfb   {\ensuremath{\mbox{\,fb}^{-1}}\xspace}

\def\mus  {\ensuremath{\rm \,\mus}\xspace}

\def\mus        {\ensuremath{\,\mu{\rm s}}\xspace}    %% microsecond
\def\to                 {\ensuremath{\rightarrow}\xspace}

\def\pep2{PEP-II}

\newcommand{\chisq}{\ensuremath{\chi^2}\xspace}

\def\gsim{{~\raise.15em\hbox{$>$}\kern-.85em
          \lower.35em\hbox{$\sim$}~}\xspace}
\def\lsim{{~\raise.15em\hbox{$<$}\kern-.85em
          \lower.35em\hbox{$\sim$}~}\xspace}

\def\CP                {\ensuremath{C\!P}\xspace}
\xspace

\newcommand{\figref}[1]{Figure~\ref{fig:#1}}
\newcommand{\tabref}[1]{Table~\ref{tab:#1}}

\def\jetset74   {\mbox{\tt Jetset \hspace{-0.5em}7.\hspace{-0.2em}4}\xspace}
\newcommand{\BABARPubYear}    {05}
\newcommand{\BABARPubNumber}  {027}

\newcommand{\SLACPubNumber} {11327}

\def\figurebox#1#2#3{%
    \def\arg{#3}%
    \ifx\arg\empty
    {\hfill\vbox{\hsize#2\hrule\hbox to #2{\vrule\hfill\vbox to #1{\hsize#2\vfill}\vrule}\hrule}\hfill}%
    \else
    {\hfill\epsfbox{#3}\hfill}%
    \fi}

\newcommand{\fisher}     {\mbox{$\cal F$}}

\newcommand{\Lzero}      {\mbox{$L_0$}}
\newcommand{\Ltwo}       {\mbox{$L_2$}}

\newcommand{\onreslumi}  {\mbox{205.4\invfb}}
\newcommand{\offreslumi} {\mbox{16.1\invfb}}
\newcommand{\bbpairs}    {\mbox{$226.0\pm2.5$ million}}
\newcommand{\nbb}        {\mbox{$N_{\BB}$}}

\newcommand{\costtb}     {\mbox{$\cos{\theta_{T}}$}}
\newcommand{\abscosttb}  {\mbox{$\left|\cos{\theta_{T}}\right|$}}

\newcommand{\half}{\mbox{${1\over2}$}}

\newcommand{\Kpppos}            {\mbox{$\Kp   \pim  \pip$}}
\newcommand{\BtoKpppos}         {\mbox{$\Bp \to \Kpppos$}}
\newcommand{\BtoKppneg}         {\mbox{$\Bm \to \Kppneg$}}
\newcommand{\Kppneg}            {\mbox{$\Km \pip \pim$}}
\newcommand{\KPP}               {\mbox{$\Kpm  \pimp \pipm$}}
\newcommand{\BtoKPP}            {\mbox{$\Bpm \to \KPP$}}

\newcommand{\fI}                 {\mbox{$f_0(980)$}}
\newcommand{\fIKp}               {\mbox{$\fI \Kp$}}
\newcommand{\fItopippim}         {\mbox{$\fI \to \pip\pim$}}
\newcommand{\fItoKpKm}           {\mbox{$\fI \to \Kp\Km$}}

\newcommand{\fII}                {\mbox{$f_2(1270)$}}
\newcommand{\fIIKp}              {\mbox{$\fII \Kp$}}
\newcommand{\fIItopippim}        {\mbox{$\fII \to \pip\pim$}}
\newcommand{\fIII}               {\mbox{$f_0(1370)$}}
\newcommand{\fIIIKp}             {\mbox{$\fIII \Kp$}}
\newcommand{\fIIItopippim}       {\mbox{$\fIII \to \pip\pim$}}

\newcommand{\fIV}                {\mbox{$f_0(1500)$}}
\newcommand{\fIVKp}              {\mbox{$\fIV \Kp$}}
\newcommand{\fIVtopippim}        {\mbox{$\fIV \to \pip\pim$}}

\newcommand{\fV}                 {\mbox{$f^{\prime}_2(1525)$}}
\newcommand{\fVKp}               {\mbox{$\fV \Kp$}}
\newcommand{\fVtopippim}         {\mbox{$\fV \to \pip\pim$}}

\newcommand{\chiczKp}            {\mbox{$\chiczero \Kp$}}
\newcommand{\chicztopippim}      {\mbox{$\chiczero \to \pip\pim$}}

\newcommand{\rhoIKp}             {\mbox{$\rhoI \Kp$}}
\newcommand{\rhoII}              {\mbox{$\rhoz(1450)$}}
\newcommand{\rhoI}               {\mbox{$\rhoz(770)$}}
\newcommand{\rhoz}               {\mbox{$\rho^0$}}
\newcommand{\rhoItopippim}       {\mbox{$\rhoI \to \pip\pim$}}
\newcommand{\rhoIIKp}             {\mbox{$\rhoII \Kp$}}
\newcommand{\rhoIItopippim}       {\mbox{$\rhoII \to \pip\pim$}}

\newcommand{\NonRes}             {\mbox{\Kpppos\ nonresonant}}

\def\Kpi   {\ensuremath{K^+\pi^-}\xspace}
\newcommand{\KstarI}            {\mbox{$\Kstarz(892)$}}
\newcommand{\KstarIpip}          {\mbox{$\KstarI \pip$}}
\newcommand{\KstarItoKppim}      {\mbox{$\KstarI \to \Kp \pim$}}
\newcommand{\KstarII}       {\mbox{$\Kstarz_0(1430)$}}

\newcommand{\KstarIIpip}         {\mbox{$\KstarII \pip$}}

\newcommand{\KpiSwave}       {\mbox{$(K\pi)^{*0}_0$}}
\newcommand{\KpiSwavepip}         {\mbox{$\KpiSwave \pip$}}
\newcommand{\KpiSwavetoKppim}     {\mbox{$\KpiSwave \to \Kp \pim$}}

\newcommand{\KstarIII}           {\mbox{$\Kstarz_{2}(1430)$}}
\newcommand{\KstarIIIpip}        {\mbox{$\KstarIII \pip$}}
\newcommand{\KstarIIItoKppim}    {\mbox{$\KstarIII \to \Kp \pim$}}
\newcommand{\KstarIV}            {\mbox{$\Kstarz(1680)$}}
\newcommand{\KstarIVpip}         {\mbox{$\KstarIV \pip$}}
\newcommand{\KstarIVtoKppim}     {\mbox{$\KstarIV \to \Kp \pim$}}

\newcommand{\Dzbpip}             {\mbox{$\Dzb \pip$}}
\newcommand{\BptoDzbpip}         {\mbox{$\Bp \to \Dzbpip$}}
\newcommand{\DzbtoKpi}           {\mbox{$\Dzb \to \Kp\pim$}}

\newcommand{\BptoKstarIpip}      {\mbox{$\Bp \to \KstarIpip$}}

\newcommand{\BptoKstarIIpip}     {\mbox{$\Bp \to \KstarIIpip$}}
\newcommand{\BptorhoIKp}         {\mbox{$\Bp \to \rhoIKp$}}
\newcommand{\BptofIKp}           {\mbox{$\Bp \to \fIKp$}}
\newcommand{\BptochiczKp}        {\mbox{$\Bp \to \chiczKp$}}

\newcommand{\JPsitoll}           {\mbox{$\jpsi \to \ellell$}}
\newcommand{\Psitoll}            {\mbox{$\psitwos \to \ellell$}}

\renewcommand{\figref}[1]{Fig.~\ref{fig:#1}}

\begin{document}

%\preprint{\babar-PUB-\BABARPubYear/\BABARPubNumber} 
%\preprint{SLAC-PUB-\SLACPubNumber} 

\begin{flushleft}
%hep-ex/0507004 \\
SLAC-PUB-\SLACPubNumber \\
\babar-PUB-\BABARPubYear/\BABARPubNumber
\end{flushleft}

\title{
{
\large \bf \boldmath Dalitz-plot analysis of the decays \BtoKPP
}
}

%author list
%% author list as of 02-Jun-2005 (635 authors)
%
\author{B.~Aubert}
\author{R.~Barate}
\author{D.~Boutigny}
\author{F.~Couderc}
\author{Y.~Karyotakis}
\author{J.~P.~Lees}
\author{V.~Poireau}
\author{V.~Tisserand}
\author{A.~Zghiche}
\affiliation{Laboratoire de Physique des Particules, F-74941 Annecy-le-Vieux, France }
\author{E.~Grauges}
\affiliation{IFAE, Universitat Autonoma de Barcelona, E-08193 Bellaterra, Barcelona, Spain }
% ...
\collaboration{The \babar\ Collaboration}
\noaffiliation

\date{
\today
\footnote{
This version of the paper corrects a coding error discovered after
submission to Phys. Rev. D and described in an Erratum submitted 26 Oct
2006.
}
}

\begin{abstract} 
\noindent

We report a Dalitz-plot analysis of the charmless hadronic decays of charged
\B\ mesons to the final state \KPP.  Using a sample of \bbpairs\ \BB\ pairs
collected by the \babar\ detector, we measure the magnitudes and phases of the
intermediate resonant and nonresonant amplitudes for both charge conjugate
decays.  We present measurements of the corresponding branching fractions and
their charge asymmetries that supersede those of previous \babar\ analyses.
We find the charge asymmetries to be consistent with zero.

\end{abstract}

\maketitle

The properties of the weak interaction, the complex quark couplings
described in the Cabibbo-Kobayashi-Maskawa matrix (CKM) elements~\cite{CKM} as
well as models of hadronic decays, can all be studied through the decay of
\B\ mesons to a three body charmless final state.
Studies of these decays can also help to clarify the nature of the resonances
involved, not all of which are well understood.
The decays \BtoKPP\ can proceed via intermediate quasi two-body resonances as
well as nonresonant decays.
The interference among these resonant and nonresonant decay modes can provide
information on the weak (\CP\ odd) and strong (\CP\ even) phases.
Theoretical predictions using various hadronic decay models exist for the
branching fractions and \CP\ asymmetries of the decays \BptoKstarIpip\ and
\BptorhoIKp~\cite{aleksan,noel,du,neubert,gronau}.
Precise measurements of the branching fractions of these modes and the
\CP\ asymmetry of \BptorhoIKp\ can discriminate among these models.
The \CP\ asymmetry of \BptoKstarIpip\ is predicted to be zero, or at least
very small, by all theoretical models.  A measurement of a large asymmetry in
this mode would therefore be a possible indication of new physics.

In this paper we present results from a full amplitude analysis for
\BtoKPP\ decay modes based on a \onreslumi\ data sample containing
\bbpairs\ \BB\ pairs (\nbb).
These data were collected with the \babar\ detector~\cite{babar} at the SLAC
\pep2\ asymmetric-energy \epem\ storage ring~\cite{pep2} operating at the
$\FourS$ resonance with center-of-mass energy of $\sqrt{s}=10.58$\gev.
An additional total integrated luminosity of \offreslumi\ was recorded
$40\mev$ below the $\FourS$ resonance and was used to study backgrounds from
continuum production.

A number of intermediate states contribute to the decay \BtoKPP.
Their individual contributions are obtained from a maximum likelihood
fit of the distribution of events in the Dalitz plot formed from the two
variables $x=m_{\Kpm\pimp}^2$ and $y=m_{\pipm\pimp}^2$.
Neglecting, at this stage, possible variations of the experimental acceptance
over the Dalitz plot, the probability density function (PDF) for signal events
is given, in the isobar formalism (see for
example~\cite{isobar1,isobar2,isobar3}), by:
\begin{equation}
P(x,y) =
\frac{\left|\sum_j c_j e^{i\theta_j} F_j(x,y)\right|^2}
{\int~{|\sum_j c_j e^{i\theta_j} F_j(x,y)|^2}~dxdy}.
\label{eq:signal-dp-prob}
\end{equation}
The amplitude for a given decay mode $j$ is $c_j e^{i\theta_j} F_j(x, y)$ with
magnitude $c_j$ and phase $\theta_j$ ($-\pi \leq \theta_j \leq \pi$). The
magnitudes and phases are measured relative to one of the contributing
channels, \KstarI\ in this analysis ($c=1$, $\theta=0$).  The distributions
$F_j$ describe the dynamics of the decay amplitudes and are a product of the
invariant mass and angular functions.
Examining the case where the resonance is formed in the $x$ variable we have:
\begin{equation}
F_j(x,y) = R_j(x) \times T_j(x,y).
\end{equation}
The $F_j$ are normalized such that:
\begin{equation}
\int \left|F_j(x,y)\right|^2 dxdy = 1.
\end{equation}
For most resonances in this analysis the $R_j$ are taken to be relativistic
Breit--Wigner lineshapes with Blatt--Weisskopf barrier
factors~\cite{blatt-weisskopf}.
There is no cut-off applied to the $R_j$ and so they are integrated over the
entire Dalitz plot.

The Breit--Wigner lineshape has the form
\begin{equation}
R_j(m) = \frac{1}{(m^2_0 - m^2) - i m_0 \Gamma(m)},
\label{eqn:BreitWigner}
\end{equation}
where $m_0$ is the nominal mass of the resonance, $m$ is the mass at which the
resonance is measured and $\Gamma(m)$ is the mass-dependent width.
In the general case of a spin $J$ resonance, the latter can be expressed as
\begin{equation}
\Gamma(m) = \Gamma_0 \left( \frac{q}{q_0}\right)^{2J+1} 
\left(\frac{m_0}{m}\right) \frac{X^2_J(q)}{X^2_J(q_0)}.
\label{eqn:resWidth}
\end{equation}
The symbol $\Gamma_0$ denotes the nominal width of the resonance.
The values of $m_0$ and $\Gamma_0$ are obtained from standard
tables~\cite{pdg2004}.
The value $q$ is the momentum of either daughter in the rest frame of the
resonance.  The symbol $q_0$ denotes the value of $q$ when $m = m_0$.
$X_J(q)$ represents the Blatt--Weisskopf barrier form
factor~\cite{blatt-weisskopf}:
\begin{eqnarray}
X_{J=0}(z) & = & 1, \\
X_{J=1}(z) & = & \sqrt{1/(1 + z^2)}, \\
X_{J=2}(z) & = & \sqrt{1/(z^4 + 3z^2 + 9)},
\label{eqn:BlattEqn}
\end{eqnarray}
where $z=rq$ and $r$, the meson radius parameter is taken to be
$4.0\gev^{-1}$~\cite{buggpc}.

For the \fI\ the Flatt\'e form~\cite{Flatte} is used.
In this case the mass-dependent width is given by the sum 
of the widths in the $\pi\pi$ and $KK$ systems:
\begin{equation}
\Gamma(m) = \Gamma_{\pi}(m) + \Gamma_{K}(m),
\label{eqn:FlatteW1}
\end{equation}
where
\begin{eqnarray}
\Gamma_{\pi}(m) &=& g_{\pi} \sqrt{m^2 - 4m^2_{\pi}} \\
\Gamma_{K}(m) &=& g_K \sqrt{m^2 - 4m^2_{K}}
\label{eqn:FlatteW2}
\end{eqnarray}
and $g_{\pi}$ and $g_K$ are the effective coupling constants squared for
\fItopippim\ and \fItoKpKm, respectively.
Below the \KpKm\ threshold the function continues analytically, the
$\Gamma_{K}$ term contributing to the real part of the denominator.

For the angular distribution terms $T_j$ we follow the Zemach tensor
formalism~\cite{Zemach1,Zemach2}.  For the case of the decay of a
spin 0 \B\ meson into a spin $J$ resonance and a spin 0 bachelor particle
this gives~\cite{asner-review}:
\begin{eqnarray}
\nonumber
T_j^{J=0} &=& 1 \\
T_j^{J=1} &=& -2 \vec{p}\cdot\vec{q} \\
T_j^{J=2} &=& {4\over3} \left[3(\vec{p}\cdot\vec{q})^2 - (|\vec{p}||\vec{q}|)^2\right],
\nonumber
\end{eqnarray}
where $\vec{p}$ is the momentum of the bachelor particle and $\vec{q}$ is the
momentum of the resonance daughter with the same charge as the bachelor
particle, both measured in the rest frame of the resonance.

The \B\ candidates are reconstructed from events that have four or more charged tracks. 
Each track is required to be well measured and to originate from the beam spot.
The \B\ candidates are formed from three-charged-track combinations and
particle identification criteria are applied to reject leptons and to separate
kaons and pions.
The average selection efficiency for kaons in our final state that have passed the
tracking requirements is about 80\% including geometrical acceptance, while
the average misidentification probability of pions as kaons is about 2\%.

Two kinematic variables are used to identify signal \B\ decays.
The first variable is $\DeltaE = E_B^* - \sqrt{s}/2$, the difference between
the reconstructed center-of-mass (CM) energy of the \B-meson candidate and
$\sqrt{s}/2$, where $\sqrt{s}$ is the total CM energy.
The second is the energy-substituted mass
$\mes = \sqrt{(s/2 + {\bf p}_i \cdot {\bf p}_\B )^2/ E_i^2 - {\bf p}^2_\B}$,
where ${\bf p}_\B$ is the \B\ momentum and ($E_i$,${\bf p}_i$) is the
four-momentum of the initial state.
The \mes\ distribution for signal events peaks near the \B\ mass with a
resolution of $2.4\mevcc$, while the \DeltaE\ distribution peaks near zero
with a resolution of $19\mev$.
Events in the interval $(-30,+60)\mev$ centered on $-4.9\mev$ in \DeltaE\ are accepted.
The shift in the central value corresponds to the mean of the signal
\DeltaE\ distribution observed in data for the channel \BptoDzbpip, \DzbtoKpi.
An asymmetric interval is chosen to reduce the amount of \BB\ background from
4-body decays.
We require events to lie in the range $5.20<\mes<5.29\gevcc$.
This range is used for an extended maximum likelihood fit to the
\mes\ distribution to determine the number of signal and background
events in our data sample.  The region is then subdivided into two areas: a
sideband region ($5.20<\mes<5.26\gevcc$) used to study the background
Dalitz-plot distribution and the signal region ($5.271<\mes<5.287\gevcc$)
where the Dalitz-plot analysis is performed.
Following the calculation of these kinematic variables the \B\ candidates
are refitted with their mass constrained to the world average value of the
\B-meson mass~\cite{pdg2004} in order to improve the Dalitz-plot position
resolution.

The dominant source of background comes from light quark and charm continuum
production ($\epem\to\qqbar$).  This background is suppressed by requirements
on event-shape variables calculated in the $\FourS$ rest frame.
For continuum background the distribution of \abscosttb, the cosine of the
angle between the thrust axis of the selected \B\ candidate and the thrust
axis of the rest of the event, is strongly peaked towards unity whereas the
distribution is uniform for signal events.
Additionally, we compute a Fisher discriminant \fisher~\cite{fisher}, a linear
combination of five variables: Legendre polynomial moments \Lzero\ and
\Ltwo~\cite{legendre}, the absolute value of the cosine of the angle between
the direction of the \B\ and the detector axis measured in the CM frame, the
absolute value of the cosine of the angle between the \B\ thrust axis and the
detector axis measured in the CM frame and the output of a multivariate
\B-flavor tagging algorithm~\cite{tagging}.
The Fisher coefficients are calculated from samples of off-resonance data and
signal Monte Carlo (MC).
The selection requirements placed on \abscosttb\ and \fisher, optimized
with MC simulated events, accept 51\% of signal events while rejecting 95\% of
background events.

Other backgrounds arise from \BB\ events. There are four main sources:
combinatorial background from three unrelated tracks;
three- and four-body \B\ decays involving an intermediate $D$ meson;
charmless two- and four-body decays with an extra or missing particle and
three-body decays with one or more particles misidentified.
We veto candidates from charm and charmonium decays with large branching
fractions by rejecting events that have invariant masses in the ranges:
$2.97 < m_{\pipm\pimp} < 3.17\gevcc$,
$3.56 < m_{\pipm\pimp} < 3.76\gevcc$,
$1.8 < m_{\Kpm\pimp} < 1.9\gevcc$ and
$1.8 < m_{\pipm\pimp} < 1.9\gevcc$.
These ranges reject decays from \JPsitoll, \Psitoll, and \DzbtoKpi\ (or \pipi)
respectively, where $\ell$ is a lepton that has been misidentified.

We study the remaining charm backgrounds that escape the vetoes and the
backgrounds from charmless \B\ decays with a large sample of MC-simulated
\BB\ decays equivalent to approximately five times the integrated luminosity
of the data sample.
The 73 decay modes that have at least one event that passes the selection
criteria are further studied with exclusive MC samples.  Of these 54 are found
to be significant, and the MC samples of these modes are used to determine the
\mes\ and Dalitz-plot distributions that are used in the likelihood fits.
These distributions are normalized to the number of predicted events in the
final data sample, which we estimate using the reconstruction efficiencies
determined from the MC, the number of \BB\ pairs in our data sample, and the
branching fractions listed by the Particle Data Group~\cite{pdg2004} and the
Heavy Flavor Averaging Group~\cite{HFAG}.
The predicted yields of \BB\ background events in the signal region are
$263\pm16$ $(270\pm16)$ for the negatively charged (positively charged) sample.

To extract the signal and \qqbar\ background fractions we perform an extended
maximum likelihood fit to the \mes\ distributions.
The signal is modeled with a double Gaussian function.
The parameters of this function are obtained from a sample of nonresonant \KPP\ MC
events and are fixed except for the mean of the core Gaussian distribution,
which is allowed to float.
The \qqbar\ \mes\ distribution is modeled with the experimentally motivated
ARGUS function~\cite{argus}. The endpoint for this ARGUS function is fixed to
$\sqrt{s}/2$ but the parameter describing the shape is left floating.
The \BB\ background \mes\ distribution is modeled as the sum of an ARGUS
function and a Gaussian distribution, whose parameters are obtained from the
\BB\ MC samples and are fixed in the fit.
The yields of signal and \qqbar\ events are allowed to float in the final fit
to the data while the yield of \BB\ background events is fixed to the value
determined above.

\begin{figure}[htb] \begin{center}
\begin{tabular}{lr}
\includegraphics[angle=0, width=\columnwidth]{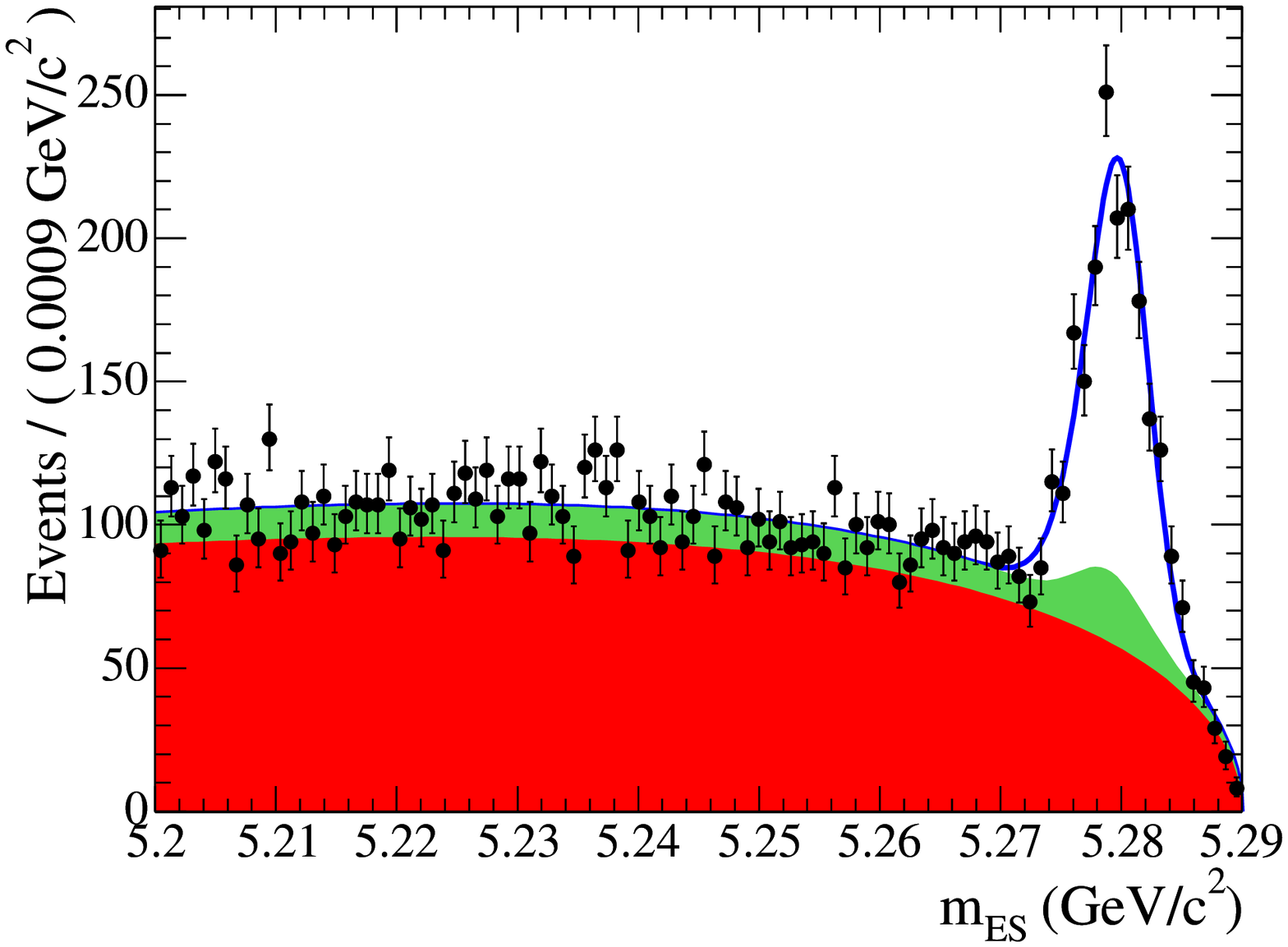}\\
\includegraphics[angle=0, width=\columnwidth]{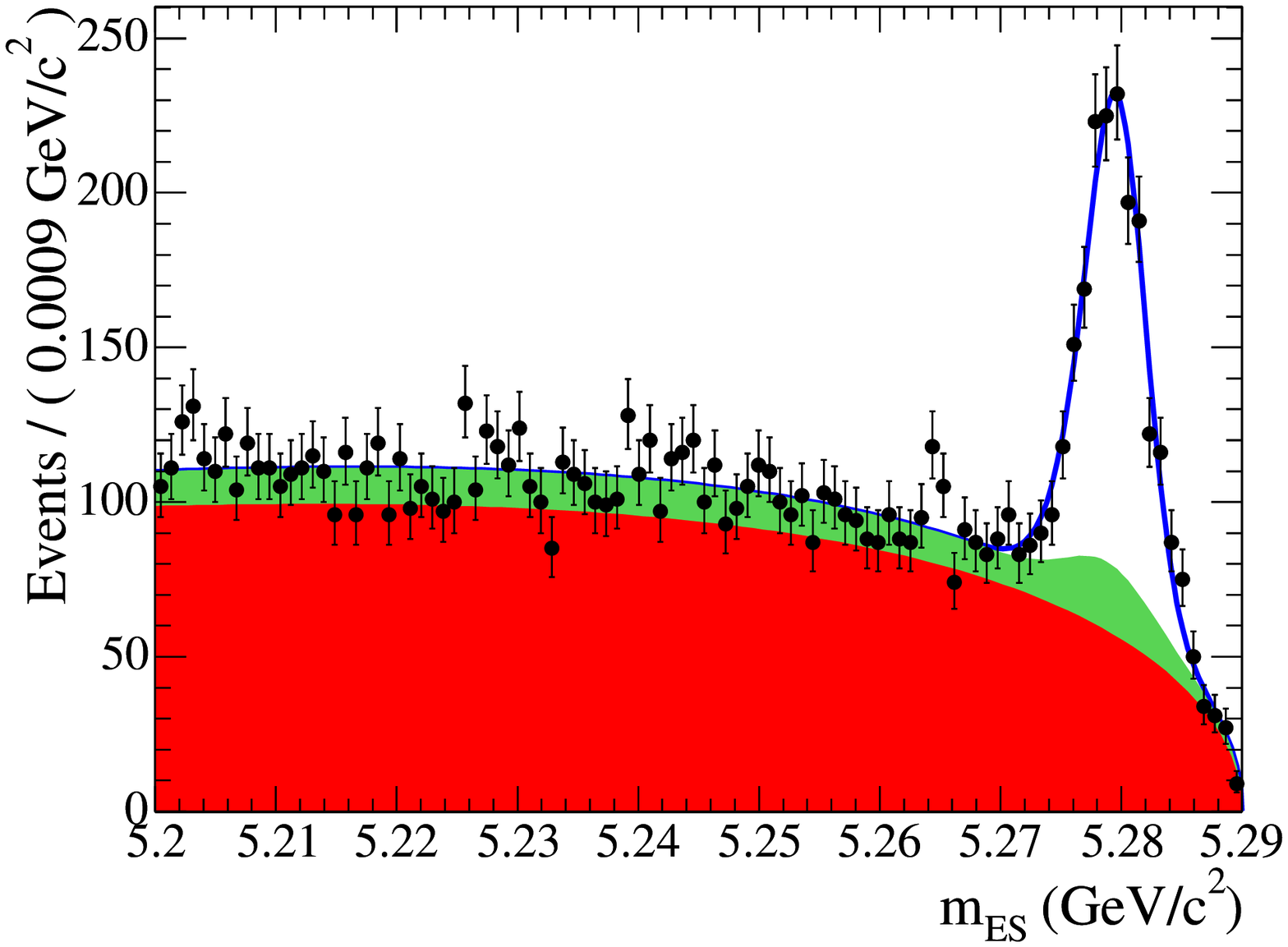}
\end{tabular}
\caption{
The \mes\ distribution, together with the fitted PDFs: 
the data are the black points with statistical error bars,
the lower solid (red/dark) area is the \qqbar\ component, 
the middle solid (green/light) area is the \BB\ background contribution, 
while the upper blue line shows the total fit result. 
The upper (lower) plot is for negatively charged (positively charged)
events.
}
\label{fig:mesfit}
\end{center} \end{figure}

The results of the fits to both the negatively charged and positively charged
samples are shown in \figref{mesfit}.
The fit yields $1047\pm56$ $(1078\pm56)$ signal events and $1016\pm25$
$(999\pm24)$ \qqbar\ events for the negative (positive) sample in the signal
region.

We independently fit the Dalitz plot of the negatively charged and positively
charged samples to extract the magnitude and phase of the intermediate
resonances and the nonresonant contribution using an unbinned
maximum-likelihood fit.  We construct a likelihood function:
\begin{eqnarray}
\label{eq:LikeEqn}
& & {\cal{L}}(x,y) = (1 - f_{q\bar{q}} - f_{B\bar{B}})\\
\nonumber
& & \frac{\left|\sum_{j=1}^{N} c_j e^{i\theta_j} F_j(x,y)\right|^2\epsilon(x,y)}
{\int~{|\sum_{j=1}^{N} c_j e^{i\theta_j} F_j(x,y)|^2\epsilon(x,y)}~dxdy}\\
\nonumber
& & +~f_{q\bar{q}}~\frac{Q(x,y)} {\int~Q(x,y)~dxdy}
    +~f_{B\bar{B}}~\frac{B(x,y)} {\int~B(x,y)~dxdy},
\end{eqnarray}
where $N$ is the number of resonant and nonresonant components in the model;
$\epsilon(x,y)$ is the signal reconstruction efficiency defined for all points
in the Dalitz plot; $Q(x,y)$ is the distribution of \qqbar\ continuum
background; $B(x,y)$ is the distribution of \BB\ background; and $f_{\qqbar}$
and $f_{\BB}$ are the fractions of \qqbar\ continuum and \BB\ background
components determined as described above and fixed in the Dalitz-plot fit.

To allow comparison among experiments we present fit fractions (FF) 
rather than amplitude magnitudes where the fit fraction is defined as the
integral of a single decay amplitude squared divided by the coherent
matrix element squared for the complete Dalitz plot:
\begin{equation}
{\it FF}_j =
\frac
{\int\left|c_j e^{i\theta_j} F_j(x,y)\right|^2 dxdy}
{\int\left|\sum_j c_j e^{i\theta_j} F_j(x,y)\right|^2 dxdy}.
\label{eq:fitfraction}
\end{equation}
The sum of all the fit fractions is not necessarily unity due to the potential
presence of net constructive or destructive interference.
The fit fraction asymmetry is defined as the difference over the sum of the
\BtoKppneg\ and \BtoKpppos\ fit fractions:
\begin{equation}
A_{\CP}^j =
\frac{\overline{\it FF}_j - {\it FF}_j}
{\overline{\it FF}_j + {\it FF}_j}.
\label{eq:fitfracasym}
\end{equation}

\begin{figure}[htb]
\begin{center}
\includegraphics[angle=0, width=\columnwidth]{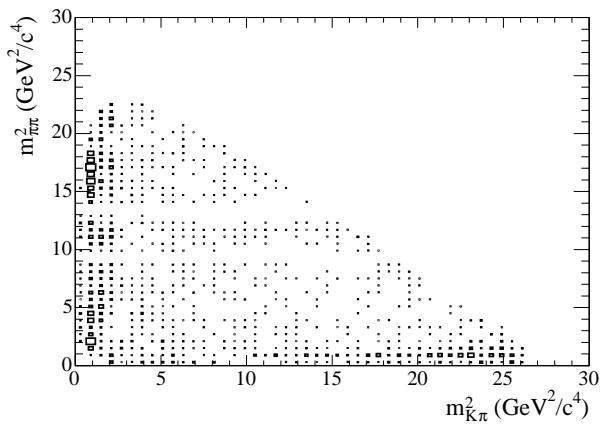}
\caption
{Background subtracted Dalitz plot of the combined \BtoKPP\ data sample in the
signal region.  The plot shows bins with greater than zero entries, the area
of the boxes being proportional to the number of entries.}
\label{fig:dp}
\end{center}
\end{figure}

The \qqbar\ continuum and \BB\ backgrounds are modeled as two-dimensional histograms
($m_{\Kpi}^2$ vs. $m_{\pipi}^2$) for the \Bp\ and \Bm\ samples with bins of size
$0.4(\gevcc)^2\times0.4(\gevcc)^2$ with linear interpolation applied between bins.
The \BB\ background distributions are taken from MC studies and the \qqbar\
distribution is taken from the \mes\ sideband data.
We expect from MC studies $1515 \pm 59$ \BB\ background events in the sideband
region ($776\pm37$ for negative events and $739\pm36$ for positive
events), which is 10.8\% of the reconstructed sideband events.  The
distribution of these events is subtracted from the \qqbar\ distribution.
In order to increase the statistical precision of this distribution we then
add off-resonance data events.  This increases the sample size by 1202 events
(605 for negative events and 597 for positive events).
The Dalitz plot of the data in the signal region after subtraction of the two
background distributions can be seen in \figref{dp}.

The two-dimensional efficiency distribution over the Dalitz plot $\epsilon(x,y)$
is calculated with 1.3 million \BtoKPP\ nonresonant MC events.  All
selection criteria are applied except for those corresponding to the invariant
mass veto regions.
The quotient is taken of two histograms, the denominator containing the true
Dalitz-plot distribution of the MC events and the numerator containing the
reconstructed MC events with corrections applied for differences between MC
and data in the particle identification and tracking efficiencies.
The efficiency shows very little variation across the majority of the Dalitz
plot but there are decreases towards the corners where one of the particles
has a low momentum.  No difference in efficiency is seen between \Bp\ and
\Bm.  The effect of experimental resolution on the signal model is neglected
since the resonances under consideration are sufficiently broad.
The average reconstruction efficiency for events in the signal box for the
nonresonant MC sample is 16.7\%.

For most resonant amplitudes the pole masses and widths are taken from the
standard Particle Data Group tables~\cite{pdg2004}.  However, there are no
consistent measurements for the coupling constants $g_{\pi}$, $g_{K}$ and the
pole mass $m_0$ of the \fI~\cite{flatte-BES,flatte-e791,flatte-wa76}.
We employ a likelihood scanning technique in order to determine the best-fit
values:
$g_{\pi}=0.11$, $g_{K}=0.36$ and $m_0=0.965\gevcc$.
The $0^+$ component of the \Kpi\ spectrum, which we denote \KpiSwave, is
poorly understood~\cite{lass,billnote,bugg}; we use the LASS
parameterization~\cite{lass,billnote} which consists of the
\KstarII\ resonance together with an effective range nonresonant component.
\begin{eqnarray}
{\cal M} &=& \frac{m_{K\pi}}{q \cot{\delta_B} - iq} \\ &+& e^{2i \delta_B} 
\frac{m_0 \Gamma_0 \frac{m_0}{q_0}}
     {(m_0^2 - m_{K\pi}^2) - i m_0 \Gamma_0 \frac{q}{m_{K\pi}} \frac{m_0}{q_0}},
\nonumber
\end{eqnarray}
where $\cot{\delta_B} = \frac{1}{aq} + \half r q$.
We have used the following values
for the scattering length and effective range parameters of this distribution:
$a=2.07\pm0.10\,(\!\gevc)^{-1}$ and $r=3.32\pm0.34\,(\!\gevc)^{-1}$~\cite{billnote}.
It has been shown in the decay $\B\to\jpsi K\pi$ that the $P-S$ phase
behaviour well matches that observed in the LASS experiment~\cite{cos2beta}.
But since this parameterization is only tested up to around 1.6\gevcc\ we
curtail the effective range term at the \Dzb\ veto.
Integrating separately the resonant part, the effective range part and the
coherent sum we find that the \KstarII\ resonance accounts for 81\%, the
effective range term 45\%, and the destructive interference between the two
terms the excess 26\% of \KpiSwave.

\begin{figure*}[htb]
\begin{center}
\includegraphics[angle=0, width=0.49\textwidth]{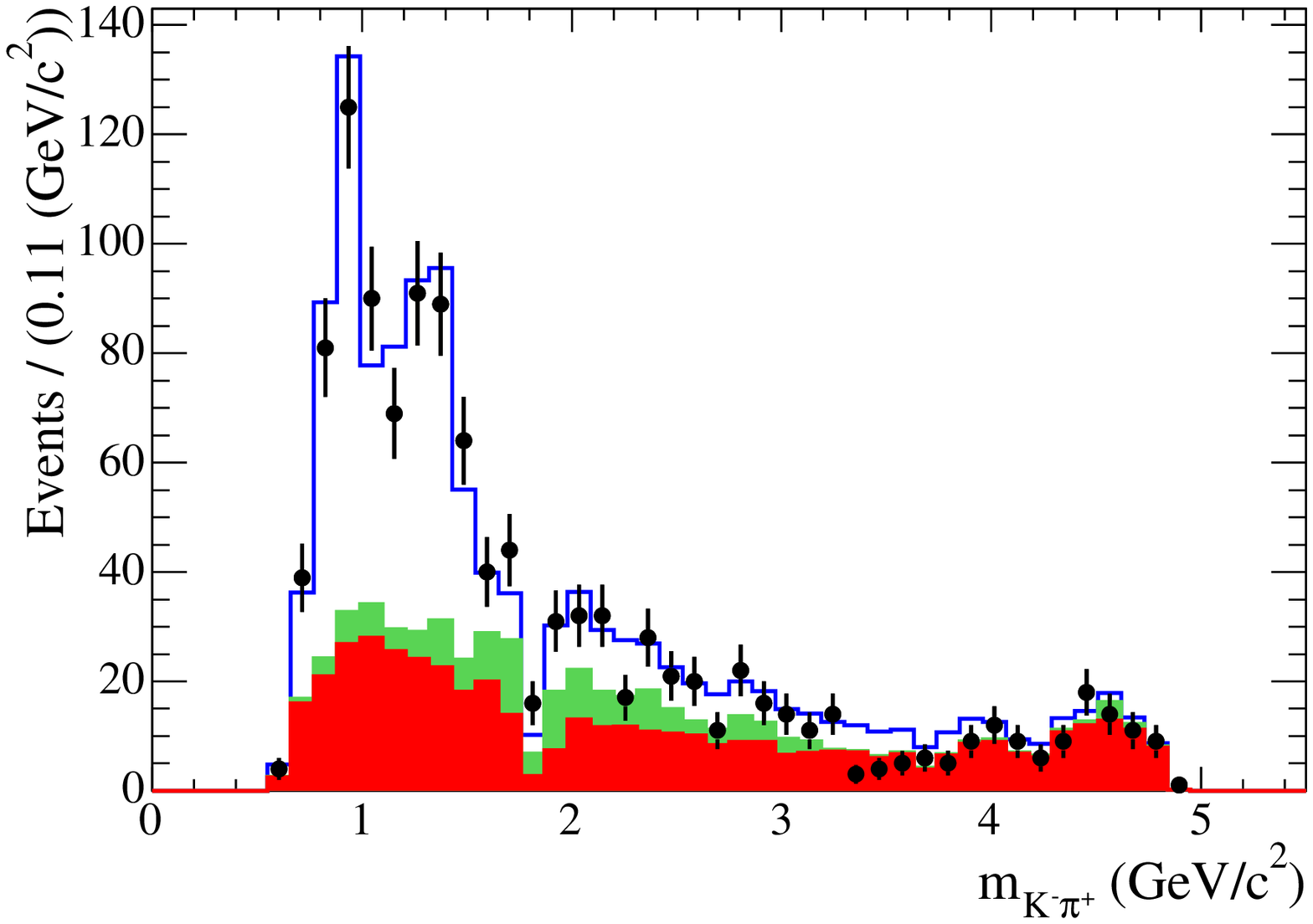}
\includegraphics[angle=0, width=0.49\textwidth]{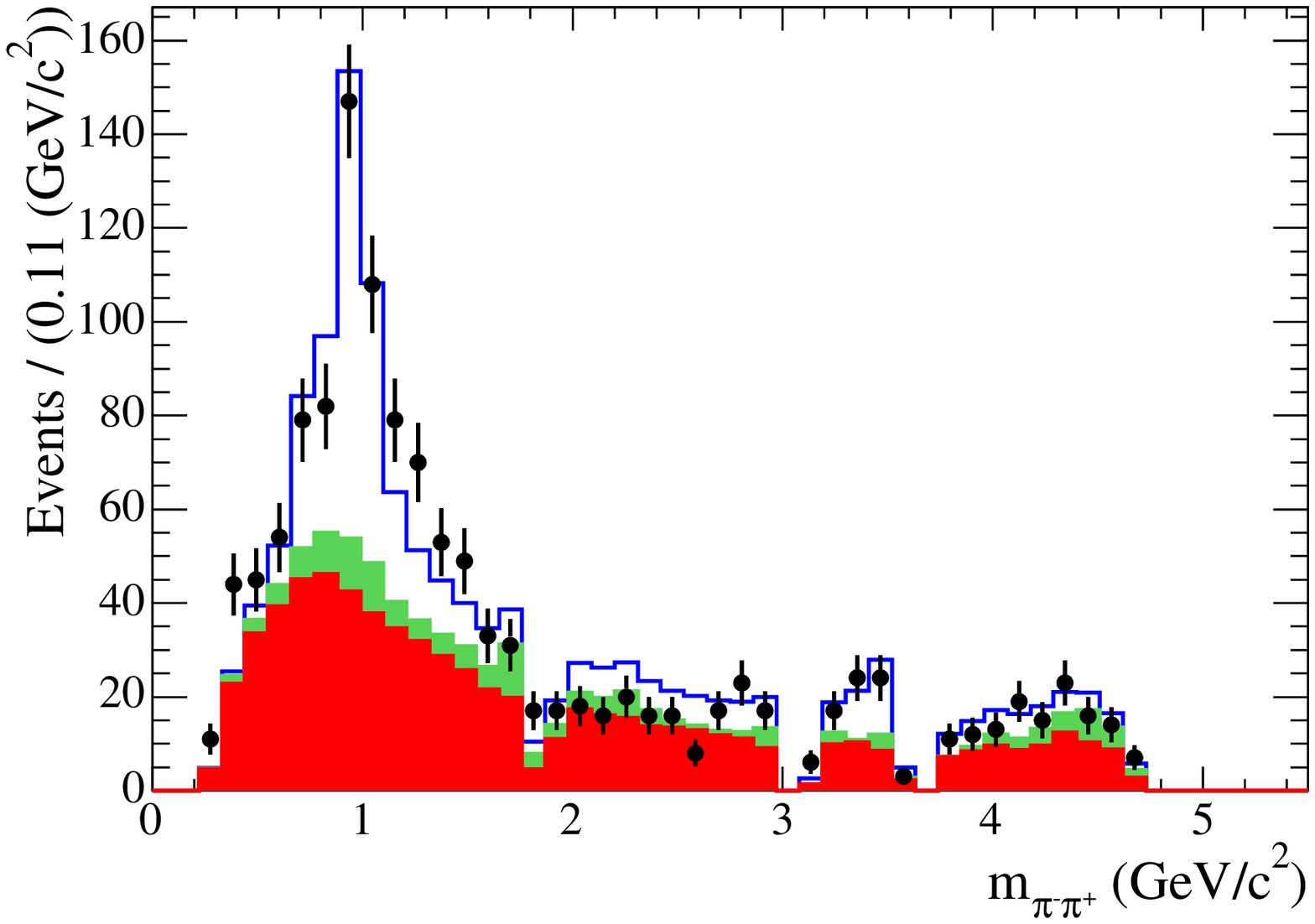}\\
\includegraphics[angle=0, width=0.49\textwidth]{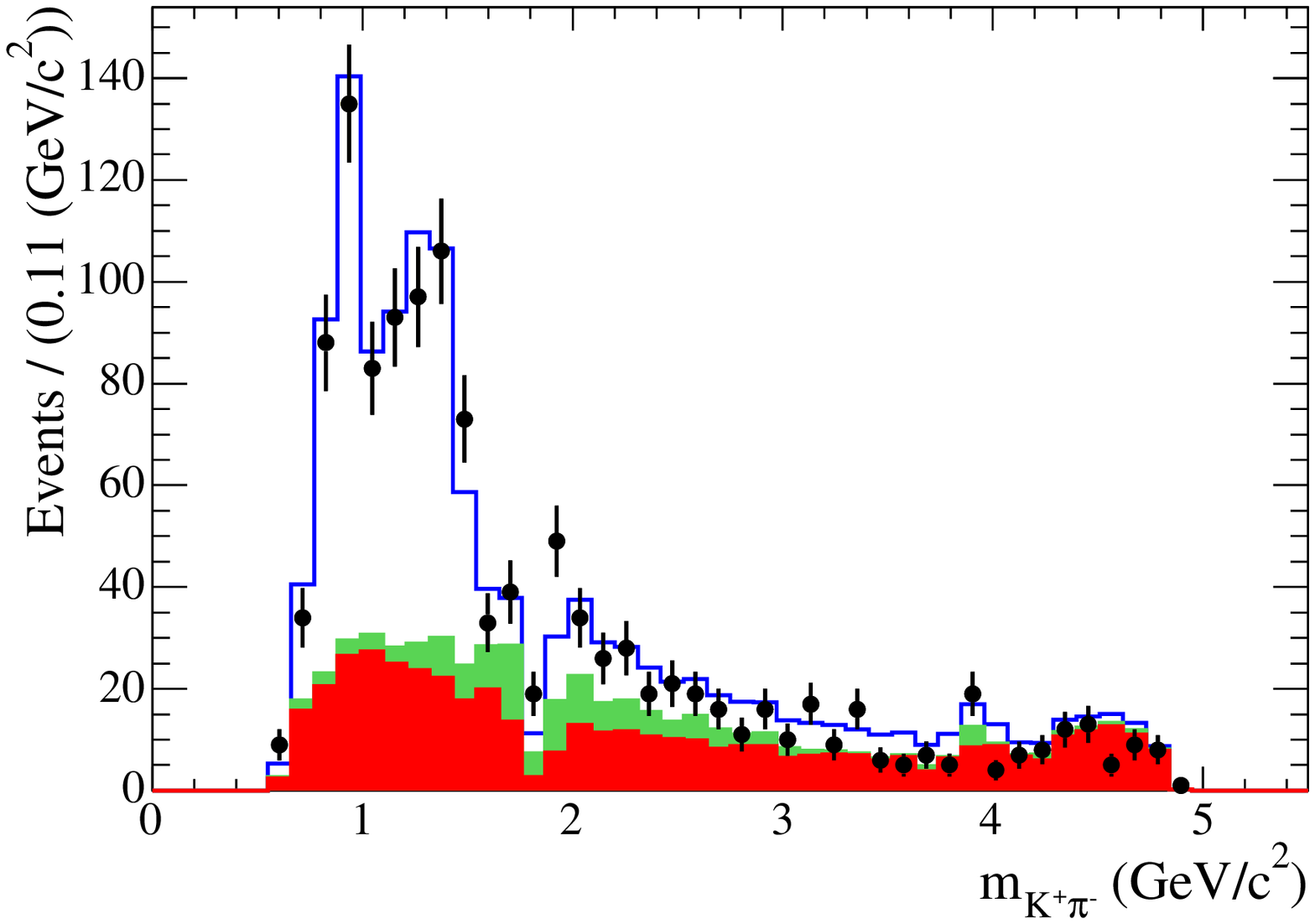}
\includegraphics[angle=0, width=0.49\textwidth]{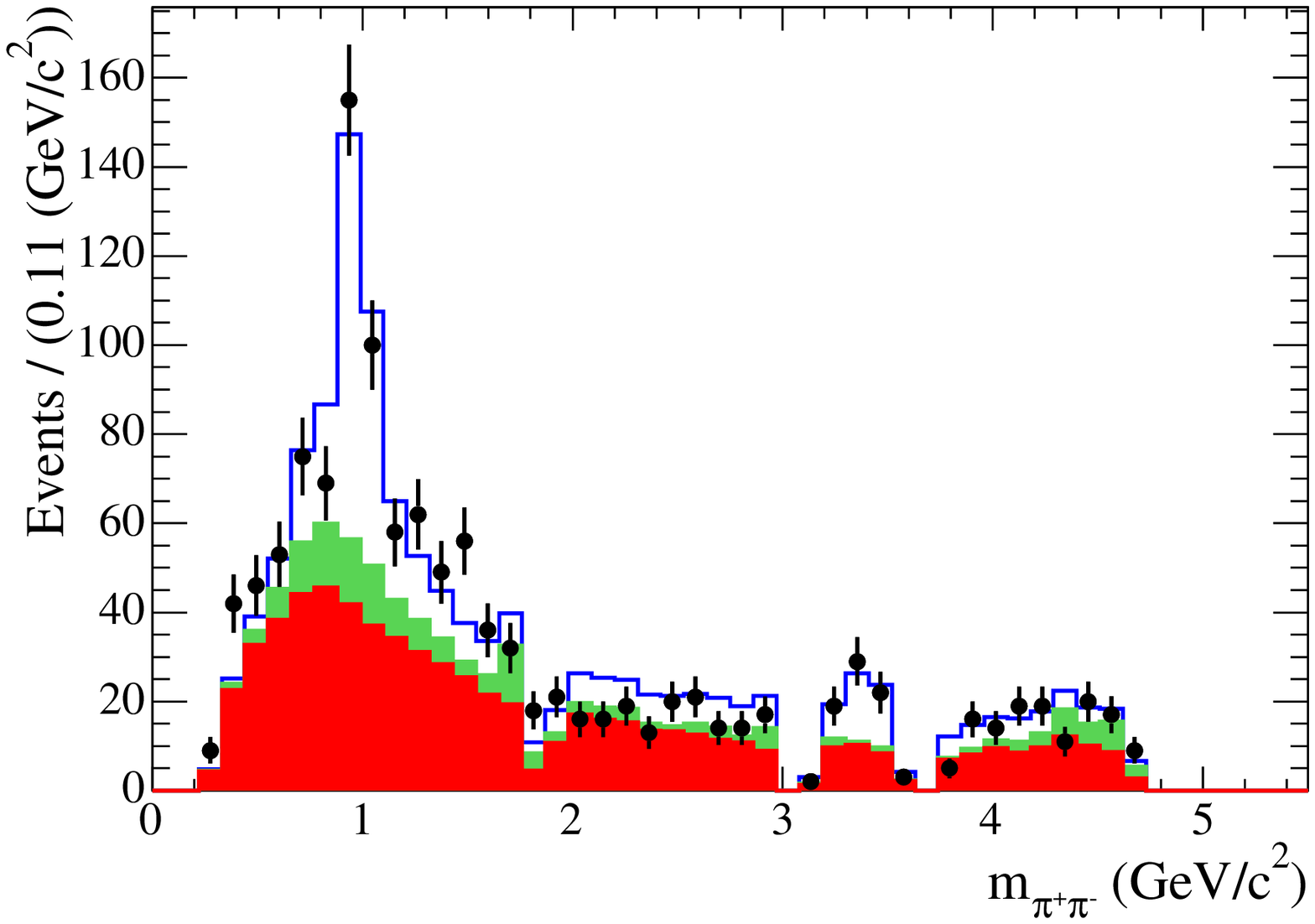}
\caption
{Invariant mass projections for the data in the signal region and the fit results.
 The upper (lower) plots are for the \Bm\ (\Bp) sample.
 The left-hand (right-hand) plots show the $m_{\Kpm\pimp}$ ($m_{\pipm\pimp}$)
 spectrum.
 The data are the black points with statistical error bars,
 the lower solid (red/dark) histogram is the \qqbar\ component, 
 the middle solid (green/light) histogram is the \BB\ background contribution, 
 while the upper blue histogram shows the total fit result. 
 The large dips in the spectra correspond to the charm vetoes.
 For the $m_{\Kpm\pimp}$ plots the requirement is made that $m_{\pipm\pimp}$
 is greater than $2\gevcc$ and vice versa.}
\label{fig:nom}
\end{center}
\end{figure*}

The nominal model comprises a phase-space nonresonant component and five
intermediate resonance states:
\KstarIpip, \KpiSwavepip, \rhoIKp, \fIKp, \chiczKp.
We choose this model using information from previous studies~\cite{KppPRD} and
the change in the goodness-of-fit observed when omitting or adding resonances.
The nonresonant component is modeled with a constant complex amplitude.
Alternative models for the nonresonant components, such as that proposed
in~\cite{belleDPpaper}, were also tested and found to make negligible
difference to the measured parameters.
The results of the fit with the nominal six-component model are shown in
\tabref{summarytable} separately for \Bm\ and \Bp\ data.
The statistical errors are calculated from MC experiments where the events are
generated from the PDFs used in the fit to data.
The \rhoI\ resonance shows the greatest difference in fit fractions between
the two samples.
The projection plots of the fit can be seen in \figref{nom}.
For the $m_{\Kpm\pimp}$ plots the requirement is made that $m_{\pipm\pimp}$ is
greater than $2\gevcc$ and vice versa in order to better illustrate the
structures present.
Using the fitted signal distribution we calculate the average reconstruction
efficiency for our signal sample to be 14.7\%.  This value, which includes all
corrections due to differences between data and MC, can be used to calculate
the inclusive branching fraction from the signal yield results.

As a measure of goodness-of-fit we evaluate the $\chi^{2}$ across the Dalitz
plot. A 75 by 75 bin histogram is used and a minimum of 10 entries per bin is
required (for those cases where this requirement is not met then neighbouring
bins are combined). The results for \Bm\ (\Bp) are a $\chi^{2}$ of 123 (148)
for a total number of 116 (121) bins and 10 free parameters in both cases. We
also calculate the same $\chi^{2}$ neglecting the region
$1.2<m_{\pipm\pimp}<1.6\gevcc$ where there is no resonant component in the
nominal fit model. The results for \Bm (\Bp) are a $\chi^{2}$ of 109 (112) for
a total number of 108 (109) bins and again 10 free parameters in both cases.

The omission of any of the nominal components from the fit results in a
significantly worse negative log-likelihood with the fitted fractions
and phases of the remaining components varying outside their error bounds.

\begin{table}[htb]
\caption{Final results of fits, with statistical, systematic and
model-dependence errors, to \Bm\ and \Bp\ data with the six component model.}
\label{tab:summarytable}
\begin{center}
\resizebox{\columnwidth}{!}{
\begin{tabular}{lcc}
\hline
Component  & \Bm\ Fit & \Bp\ Fit \\
\hline
\hline
\vspace{2mm}
\KstarI\ Fraction (\%)     & $15.0\pm1.6\pm0.6^{+0.5}_{-1.3}$          & $13.1\pm1.5\pm0.6^{+0.7}_{-1.2}$          \\
\vspace{2mm}
\KstarI\ Magnitude         & $1.0$ FIXED                               & $1.0$ FIXED                               \\
\vspace{1mm}
\KstarI\ Phase             & $0.0$ FIXED                               & $0.0$ FIXED                               \\
\hline
\vspace{2mm}
\KpiSwave\ Fraction(\%)   & $49.6\pm2.4\pm0.7^{+0.8}_{-5.0}$           & $56.4\pm2.4\pm0.7^{+3.3}_{-5.2}$          \\
\vspace{2mm}
\KpiSwave\ Magnitude       & $1.82\pm0.12\pm0.04^{+0.07}_{-0.09}$      & $2.08\pm0.15\pm0.06^{+0.11}_{-0.14}$      \\
\vspace{1mm}
\KpiSwave\ Phase           & $-0.38\pm0.12\pm0.03^{+0.08}_{-0.05}$     & $0.01\pm0.12\pm0.03\pm0.06$               \\
\hline
\vspace{2mm}
\rhoI\ Fraction (\%)       & $10.5\pm1.7\pm0.4^{+0.6}_{-2.5}$          & $5.4\pm1.3\pm0.5^{+0.9}_{-1.4}$           \\
\vspace{2mm}
\rhoI\ Magnitude           & $0.837\pm0.079\pm0.031^{+0.016}_{-0.076}$ & $0.642\pm0.092\pm0.042^{+0.060}_{-0.076}$ \\
\vspace{1mm}
\rhoI\ Phase               & $-0.55\pm0.38\pm0.08^{+0.75}_{-0.65}$     & $1.19\pm0.62\pm0.16^{+0.95}_{-0.57}$      \\
\hline
\vspace{2mm}
\fI\ Fraction (\%)         & $16.1\pm2.1\pm0.4^{+1.1}_{-2.1}$          & $13.5\pm1.9\pm0.4^{+0.7}_{-2.5}$          \\
\vspace{2mm}
\fI\ Magnitude             & $1.037\pm0.080\pm0.026^{+0.053}_{-0.072}$ & $1.015\pm0.095\pm0.036^{+0.059}_{-0.092}$ \\
\vspace{1mm}
\fI\ Phase                  & $1.23\pm0.34\pm0.07^{+0.69}_{-0.52}$     & $2.29\pm0.56\pm0.13^{+1.39}_{-0.56}$      \\
\hline
\vspace{2mm}
\chiczero\ Fraction (\%)   & $0.84\pm0.44\pm0.14^{+0.07}_{-0.08}$      & $1.20\pm0.50\pm0.13^{+0.07}_{-0.08}$      \\
\vspace{2mm}
\chiczero\ Magnitude       & $0.237\pm0.047\pm0.009^{+0.018}_{-0.013}$ & $0.302\pm0.052\pm0.016^{+0.016}_{-0.016}$ \\
\vspace{1mm}
\chiczero\ Phase           &  $2.55\pm0.40\pm0.09^{+0.35}_{-0.11}$     & $-2.52\pm0.37\pm0.35^{+0.18}_{-0.17}$     \\
\hline
\vspace{2mm}
Nonresonant Fraction (\%)  & $4.3\pm1.3\pm0.8^{+1.3}_{-1.4}$           & $4.6\pm1.5\pm0.9^{+1.8}_{-0.4}$           \\
\vspace{2mm}
Nonresonant Magnitude      & $0.54\pm0.07\pm0.06^{+0.10}_{-0.10}$      & $0.60\pm0.11\pm0.06^{+0.09}_{-0.01}$      \\
\vspace{1mm}
Nonresonant Phase          & $-2.29\pm0.33\pm0.09^{+0.61}_{-0.40}$     & $-1.85\pm0.28\pm0.07^{+0.38}_{-0.26}$     \\
\hline
\end{tabular}
}
\end{center}
\end{table}

We have tested for the sensitivity of these results to additional
resonances that can be added in the fit function.
In the $\pi\pi$ spectrum there are possible
higher resonances including \fII, \fIII, \rhoII, \fIV and \fV; in the $K\pi$
spectrum there are possible \KstarIII\ and \KstarIV\ resonances.  Each of
these resonances is added in turn to the nominal signal model, to form an
extended model, and the Dalitz-plot fit is repeated.
In general, adding another component does not significantly affect the
measured fit fractions and phases of the six nominal components.
We place upper limits on each of the possible additional components
(\tabref{conclusion-results-summary}).

The systematic uncertainties that affect the measurement of the fit fractions and
phases are evaluated separately for \Bm\ and \Bp.
Each bin of the efficiency, \qqbar\ and \BB\ background histograms is
fluctuated independently in accordance with its errors and the nominal fit
repeated.  The fractions of \BB\ and \qqbar\ events are varied in accordance
with their errors and the fits repeated.
To confirm the fitting procedure, 500 MC experiments were performed in which the
events are generated from the PDFs used in the fit to data.
A small fit bias is observed for the \chiczero\ and nonresonant components and is
included in the systematic uncertainties.
There is a contribution to the fit fraction asymmetry from possible detector
charge bias, which has been estimated in previous studies to be
1\%~\cite{johnprl}. A 0.4\% systematic error on the total branching fraction
comes from the error on the predicted number of \BB\ background events.
The systematic uncertainties for particle identification and tracking
efficiency corrections are 4.2\% and 2.4\% respectively.
The calculation of \nbb\ has a total uncertainty of 1.1\%.
The efficiency corrections due to the selection requirements on \costtb, the
Fisher discriminant, \DeltaE\ and \mes\ have also been calculated from
\BptoDzbpip, \DzbtoKpi\ data and MC samples, and the error on these corrections
is incorporated into the branching fraction systematic uncertainties.

In addition to the above systematic uncertainties we also calculate a
model-dependence uncertainty that characterizes the uncertainty on the
results due to elements of the signal Dalitz-plot model.
The first of these elements consists of the two parameters of the LASS model
of the $K\pi$ S-wave and is calculated by refitting adjusting the parameters
of the LASS model within their experimental errors.
The second consists of the three parameters of the \fI\ resonance and is
evaluated by refitting with the parameter values measured by the BES
collaboration~\cite{flatte-BES}.
The third element is due to the different possible models for the nonresonant
component and is evaluated by refitting with the parameterization proposed by
Belle~\cite{belleDPpaper}.
The fourth element is the uncertainty due to the composition of
the signal model and reflects observed changes in the parameters of the
nominal components when the data are fitted with the extended models.
The uncertainties from each of these elements are added in quadrature to give
the final model-dependence uncertainty.

\begin{table*}[htb]
\caption
{Summary of measurements of branching fractions (averaged over charge
conjugate states) and \CP\ asymmetries.
The first error is statistical, the second is systematic and the third
represents the model dependence.
}
\label{tab:conclusion-results-summary}
\begin{center}
\begin{tabular}{lccc}
\hline
Mode & $\BR(\Bp \to {\rm Mode}) (10^{-6})$ & 90\% CL UL $(10^{-6})$ & $A_{\CP}$ (\%) \\
\hline
\Kpppos\ Total & $64.1\pm2.4\pm4.0$ & $-$ & $-1.3\pm3.7\pm1.1$ \\
\hline
\KstarIpip; \KstarItoKppim     & $8.99\pm0.78\pm0.48^{+0.28}_{-0.39}$ & $-$     & $6.8\pm7.8\pm5.7^{+4.0}_{-3.5}$  \\
\KpiSwavepip; \KpiSwavetoKppim & $34.0\pm1.7\pm1.5^{+1.2}_{-1.6}$     & $-$     & $-6.4\pm3.2\pm2.0^{+1.1}_{-1.7}$ \\
\rhoIKp; \rhoItopippim         & $5.07\pm0.75\pm0.35^{+0.42}_{-0.68}$ & $-$     & $32\pm13\pm6^{+8}_{-5}$          \\
\fIKp; \fItopippim             & $9.47\pm0.97\pm0.46^{+0.42}_{-0.75}$ & $-$     & $8.8\pm9.5\pm2.6^{+9.3}_{-5.0}$  \\
\chiczKp; \chicztopippim       & $0.66\pm0.22\pm0.07\pm0.03$          & $<1.1$  & $-$ \\
\NonRes                        & $2.85\pm0.64\pm0.41^{+0.70}_{-0.34}$ & $<6.5$  & $-$ \\
\KstarIIIpip; \KstarIIItoKppim & $-$                                  & $<7.7$  & $-$ \\
\KstarIVpip; \KstarIVtoKppim   & $-$                                  & $<3.8$  & $-$ \\
\fIIKp; \fIItopippim           & $-$                                  & $<8.9$  & $-$ \\
\fIIIKp; \fIIItopippim         & $-$                                  & $<10.7$ & $-$ \\
\rhoIIKp; \rhoIItopippim       & $-$                                  & $<11.7$ & $-$ \\
\fIVKp; \fIVtopippim           & $-$                                  & $<4.4$  & $-$ \\
\fVKp; \fVtopippim             & $-$                                  & $<3.4$  & $-$ \\
\hline
\end{tabular}
\end{center}
\end{table*}

In order to make comparisons with previous measurements and predictions from
factorization models we multiply each fit fraction by the total branching
fraction to calculate the branching fraction of the mode.  These branching
fractions from each of the charge-separated fits are then averaged.
For components that do not have statistically significant branching fractions
90\% confidence level upper limits are determined.  Upper limits are also
calculated for the components added in the extended signal models.
Upper limits are calculated from MC experiments where each experiment is
generated from the fitted PDFs but with all sources of systematic uncertainty
varied in accordance with their errors.
The measured branching fractions, averaged over charge-conjugate states, and
\CP\ asymmetries from this analysis are summarized in
\tabref{conclusion-results-summary}.
Charge conjugates are included implicitly throughout this table and the
following discussion.

The total \BtoKPP\ branching fraction ($(64.1\pm2.4\pm4.0)\times10^{-6}$) has
been measured with increased statistical accuracy and is compatible with
previous \babar\ measurements.  It differs from Belle's measurement of
$(46.6\pm2.1\pm4.3)\times10^{-6}$~\cite{belleDPpaper}, which is significantly
smaller, even after accounting for the \chiczKp\ mode, which Belle does not
include.
This result was cross checked by using the same procedure to measure the
\BptoDzbpip; \DzbtoKpi\ branching fraction, which was found to be consistent
with the PDG value~\cite{pdg2004}.
The total charge asymmetry has been measured to be consistent with zero to a
higher degree of precision than previous measurements.

After correcting for the secondary branching fraction
$\BR(\KstarItoKppim)={2\over3}$ we find the \BptoKstarIpip\ branching fraction
to be:
$(13.5\pm1.2\pm0.7^{+0.4}_{-0.6})\times10^{-6}$.
This is smaller than that measured in previous analyses that do not perform an
amplitude fit to the Dalitz plot~\cite{KppPRD,belleRegpaper} but slightly
larger than the value reported by Belle in their amplitude
analysis~\cite{belleDPpaper}.
The branching fraction measurement of \BptorhoIKp\ is the first measurement of
the mode from \babar\ and is consistent with that from
Belle~\cite{belleDPpaper}.  It is also broadly compatible with many
theoretical predictions~\cite{aleksan,noel,du,neubert,gronau}.
The \BptofIKp\ branching fraction is in good agreement with earlier
analyses~\cite{KppPRD,belleRegpaper} and the recent Belle amplitude
analysis~\cite{belleDPpaper}.
The forward-backward asymmetry apparent in both the \KstarI\ and \fI\ bands in
\figref{dp} is well reproduced by the fit and is not due to reconstruction
efficiency effects but to $S-P$ interference in the Dalitz plot.

The \KpiSwave\ component appears to be well modeled by the LASS
parameterization, which consists of a nonresonant effective range term plus a
relativistic Breit--Wigner term for the \KstarII\ resonance itself.
Removing the phase-space nonresonant component from the nominal model gives
very little change in the goodness-of-fit \chisq\ or the fit likelihood.
It is unclear whether this component is required in addition to the
nonresonant part of the \KpiSwave\ component.
We can calculate the branching fraction for \BptoKstarIIpip\ using our
knowledge of the composition of the \KpiSwave\ component and find it to be:
$(44.4\pm2.2\pm2.0^{+1.6}_{-2.1}\pm4.5)\times10^{-6}$,
where the fourth error is due to the uncertainty on the branching fraction of
$\KstarII \to K\pi$ combined with the uncertainty on the proportion of the
\KpiSwave\ component due to the \KstarII\ resonance.
The Belle collaboration finds a similarly large \KstarII\ branching fraction
though they treat the \KstarII\ as a separate component from the rest of the
S-wave, modeled as a nonresonant component that has variation in magnitude but
no variation in phase over the Dalitz plot~\cite{belleDPpaper}.

For \BptorhoIKp\ and \BptochiczKp\ the differences in phase between the \Bp\ and
\Bm\ decays are $1.74\pm0.73$ (2.4 standard deviations, $\sigma$) and
$1.21\pm0.54$ ($2.2\sigma$) respectively, where the errors are statistical
only.
The \KstarIpip\ charge asymmetry is consistent with zero, as expected from the
Standard Model predictions~\cite{aleksan,noel,du,neubert,gronau}.
There is no evidence of new physics entering the penguin diagram loop.

We are grateful for the excellent luminosity and machine conditions
provided by our \pep2\ colleagues, 
and for the substantial dedicated effort from
the computing organizations that support \babar.
The collaborating institutions wish to thank 
SLAC for its support and kind hospitality. 
This work is supported by
DOE
and NSF (USA),
NSERC (Canada),
IHEP (China),
CEA and
CNRS-IN2P3
(France),
BMBF and DFG
(Germany),
INFN (Italy),
FOM (The Netherlands),
NFR (Norway),
MIST (Russia), and
PPARC (United Kingdom). 
Individuals have received support from CONACyT (Mexico), A.~P.~Sloan Foundation, 
Research Corporation,
and Alexander von Humboldt Foundation.
\bibliography{references}
\bibliographystyle{apsrev}

\end{document}